\documentclass[aps,prd,twocolumn,superscriptaddress,amsmath,amssymb]{revtex4-1}

\usepackage{hyperref}
\usepackage{graphicx}

\newcommand{\etal}{{\it et al.}}
\newcommand{\unit}[1]{~\mathrm{#1}}
\newcommand{\shiki}[1]{Eq.~(\ref{#1})}
\newcommand{\zu}[1]{Fig.~{\ref{#1}}}
\newcommand{\hyou}[1]{Table~{\ref{#1}}}
\newcommand{\Reference}[1]{Ref.~{\cite{#1}}}
\newcommand{\mneutron}{m_{\rm n}}

\begin{document}

\title{Ultralight vector dark matter search with\\ auxiliary length channels of gravitational wave detectors}

\author{Yuta Michimura}
  \email{michimura@phys.s.u-tokyo.ac.jp}
  \affiliation{Department of Physics, University of Tokyo, Bunkyo, Tokyo 113-0033, Japan}
\author{Tomohiro Fujita}
  \affiliation{Institute for Cosmic Ray Research, University of Tokyo, Kashiwa, Chiba 277-8582, Japan}
\author{Soichiro Morisaki}
  \affiliation{Institute for Cosmic Ray Research, University of Tokyo, Kashiwa, Chiba 277-8582, Japan}
\author{Hiromasa Nakatsuka}
  \affiliation{Institute for Cosmic Ray Research, University of Tokyo, Kashiwa, Chiba 277-8582, Japan}
\author{Ippei Obata}
  \affiliation{Department of Physics, Kyoto University, Sakyo, Kyoto 606-8502, Japan}
  \affiliation{Max-Planck-Institut f{\"u}r Astrophysik, Karl-Schwarzschild-Str. 1, 85741 Garching, Germany}
\date{\today}

\begin{abstract}
Recently, a considerable amount of attention has been given to the search for ultralight dark matter by measuring the oscillating length changes in the arm cavities of gravitational wave detectors. Although gravitational wave detectors are extremely sensitive for measuring the differential arm length changes, the sensitivity to dark matter is largely attenuated, as the effect of dark matter is mostly common to arm cavity test masses. Here, we propose to use auxiliary length channels, which measure the changes in the power and signal recycling cavity lengths and the differential Michelson interferometer length. The sensitivity to dark matter can be enhanced by exploiting the fact that auxiliary interferometers are more asymmetric than two arm cavities. We show that the sensitivity to $U(1)_{B-L}$ gauge boson dark matter with masses below $7\times 10^{-14}$~eV can be greatly enhanced when our method is applied to a cryogenic gravitational wave detector KAGRA, which employs sapphire test masses and fused silica auxiliary mirrors. We show that KAGRA can probe more than an order of magnitude of unexplored parameter space at masses around $1.5 \times 10^{-14}$~eV, without any modifications to the existing interferometer.
\end{abstract}

\maketitle

\section{Introduction}
Despite strong observational evidence for the existence of dark matter, its identity and properties remain a mystery.
For decades, conventional dark matter searches have focused on weakly interacting massive particles (WIMPs) with masses around the weak scale. However, no evidence for WIMPs suggests the importance of testing other dark matter candidates spreading over a large target mass range of $10^{-22} \unit{eV} \lesssim m \lesssim 10^{69} \unit{eV}$, extending over 90 orders of magnitude~\cite{NewEra}.
Among various candidates, bosonic ultralight fields with masses of $10^{-22} \unit{eV} \lesssim m \lesssim 1 \unit{eV}$ are well motivated by cosmology because they behave as classical wave fields, rather than individual particles~\cite{ULDMWitten,ULDMFerreira}.

Recently, a number of novel ideas have been proposed to search for ultralight dark matter candidates using laser interferometers at various scales~\cite{AxionInterferometry,DANCE,ADBC,LinearAxion,QuantumAxion,Stadnik2015,Stadnik2016,Geraci2019,Grote2019,Arvanitaki2015,Morisaki2019,Graham2016,Pierce2018,Carney2019,Manley2020}, from centimeter-scale optical cavities to kilometer-scale gravitational wave detectors such as Advanced LIGO (aLIGO)~\cite{aLIGO,aLIGOO1NB}, Advanced Virgo~\cite{AdV}, and KAGRA~\cite{AsoKAGRA,bKAGRAPhase1}.
Laser interferometers are very sensitive to oscillating changes in the phase velocity of photons or the optical path length and are thus suitable devices to look for ultralight dark matter candidates that cause these effects.
Axion-like particles can be searched for by measuring the phase velocity difference between the left- and right-handed polarized photons~\cite{AxionInterferometry,DANCE,ADBC,LinearAxion,QuantumAxion}.
Scalar fields that cause time variation of the fine structure constant or the particle masses can be probed by measuring the size changes in mirrors or spacers of rigid optical cavities~\cite{Stadnik2015,Stadnik2016,Geraci2019,Grote2019}.
These fields can also be searched for by measuring the acceleration caused by the spatial gradient of the mass of the mirrors~\cite{Arvanitaki2015,Morisaki2019}.

The massive vector field weakly coupled to the standard model sector (also known as dark photon) via the baryon number, $B$, or the baryon minus lepton number, $B-L$, has increasingly received attention as yet another ultralight dark matter candidate.
The theoretical attempts to identify $U(1)_{B}$ or $U(1)_{B-L}$ as gauge symmetry have been explored as a natural extension of the standard model. Although $U(1)_{B}$ symmetry is anomalous in the standard model, the anomaly can be canceled by introducing an additional degree of freedom, for example, the Green-Schwarz mechanism~\cite{Green:1984sg}.
In this case, the $U(1)_B$ gauge boson acquires its mass through the Stueckelberg mechanism.
On the other hand, $U(1)_{B-L}$ is anomaly free, and can be gauged without additional ingredients.
The $U(1)_{B-L}$ gauge boson can also acquire its mass via the Higgs mechanism. The mass is then proportional to the gauge coupling constant.
Therefore, if the $U(1)_{B-L}$ gauge boson is an ultralight dark matter,
we expect that its coupling to the standard model is suppressed,
and requires highly sensitive experiments to detect it.
Several proposals have been made to search for such vector fields by measuring the oscillating forces acting on mirrors with laser interferometers~\cite{Graham2016,Pierce2018,Carney2019,Manley2020}.

Among these proposals to probe various ultralight dark matter candidates, the use of gravitational wave detectors is often considered, owing to their extremely high displacement sensitivity, on the order of $10^{-20} \unit{m/\sqrt{Hz}}$ at around 100 Hz~\cite{aLIGOO1NB}.
Although gravitational wave detectors are highly sensitive to measure differential length changes in two perpendicular arm cavities, length changes driven by ultralight scalar or vector fields are mostly common to arm cavity test masses, and most of the effects are canceled out.
The sensitivity to dark matter couplings therefore relies on a slight asymmetry between the arms or slight difference in the phase of the dark matter field at two distant test masses of the arm cavity.

In this paper, we propose the use of auxiliary length channels of gravitational wave detectors, such as the channels to monitor changes in the power and signal recycling cavity lengths and differential Michelson interferometer length, to enhance the sensitivity to dark matter couplings.
We especially consider vector fields and show that the sensitivity can be improved compared to the search using the main differential arm length channel, when the main test masses and auxiliary mirrors have different charges to which the dark matter is coupled.
This condition is satisfied for the coupling between the $U(1)_{B-L}$ gauge field and cryogenic gravitational wave detectors such as KAGRA, which employ sapphire test masses and fused silica auxiliary mirrors.

In the following, we start by briefly introducing the interferometer configuration of KAGRA and aLIGO, and define the length channels of the interferometer. Then, we show how these channels are modulated by the coupling of the $U(1)_{B}$ and $U(1)_{B-L}$ gauge fields to the associated charge of the mirrors. Next, we describe the prospected sensitivity curves for the coupling constant for each length channel. Finally, we provide a brief discussion and conclude our results. Throughout this paper, we use natural units $\hbar=c=\epsilon_0=1$.

\section{Length channels of gravitational wave detectors}

The interferometer configuration of KAGRA and aLIGO is a dual-recycled Fabry-P{\'e}rot-Michelson interferometer, as shown in \zu{Config}.
It is based on a Michelson interferometer that has two Fabry-P{\'e}rot cavities of length $L_{\rm x}=L_{\rm y} \equiv L_{\rm arm}$ in perpendicular arms. The arm cavity is formed by the input test mass (ITM) and the end test mass (ETM), and the main gravitational wave signal is imprinted in the differential arm length (DARM).
The differential Michelson interferometer length (MICH) between the beam splitter (BS) and two ITMs are controlled at the dark fringe at the anti-symmetric port where the DARM channel is obtained.
Most of the input beam is reflected in the direction of the laser source, where a power recycling mirror (PRM) is located.
The PRM and two ITMs form a power recycling cavity, and its length, power recycling cavity length (PRCL), is controlled to effectively enhance the input power.
Additionally, a signal recycling mirror (SRM) is located at the anti-symmetric port to change the frequency response of the interferometer by tuning the signal recycling cavity length (SRCL).
The length changes of these auxiliary degrees of freedom can be obtained from the reflection port and the pick-off port of the power recycling cavity.
Using the length symbols in \zu{Config}, changes in DARM, MICH, PRCL and SRCL can be written as
\begin{eqnarray}
 \delta L_{\rm DARM} &=& \delta  (L_{\rm x} - L_{\rm y}), \\ \label{eq:darm}
 \delta L_{\rm MICH} &=& \delta (l_{\rm x} - l_{\rm y}), \\ \label{eq:mich}
 \delta L_{\rm PRCL} &=& \delta [(l_{\rm x} + l_{\rm y})/2 + l_{\rm p}], \\  \label{eq:prcl}
 \delta L_{\rm SRCL} &=& \delta [(l_{\rm x} + l_{\rm y})/2 + l_{\rm s}],  \label{eq:srcl}
\end{eqnarray}
respectively. Here, $l_{\rm p}$ ($l_{\rm s}$) is the folded optical path length between PRM (SRM) and BS. The interferometer length parameters are listed in \hyou{Lengths}.

\begin{figure}
\begin{center}
\includegraphics[width=\hsize]{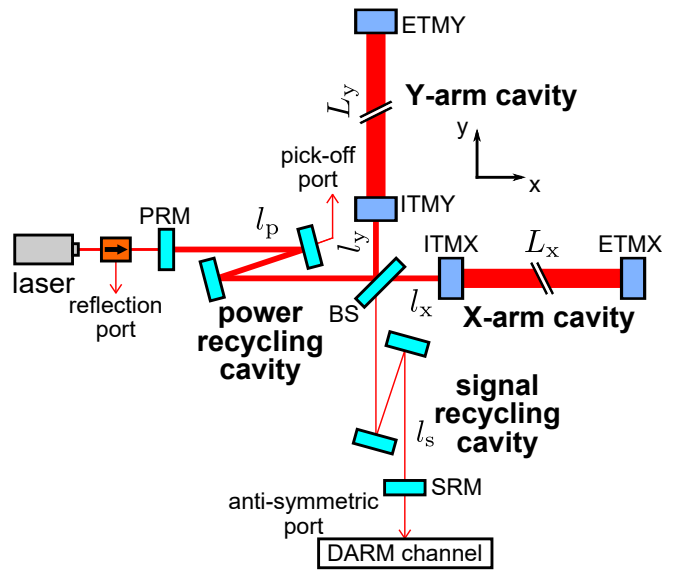}
\end{center}
\caption{\label{Config} The schematic of a dual-recycled Fabry-P{\'e}rot-Michelson interferometer, such as KAGRA and aLIGO. ITM (ETM): input (end) test mass, BS: beam splitter, PRM: power recycling mirror, SRM: signal recycling mirror.}
\end{figure}

\begin{table}
    \caption{\label{Lengths} KAGRA interferometer length parameters used for the sensitivity calculation~\cite{AsoKAGRA}. See \zu{Config} for the length symbols. Length $l^{\prime}_{\rm p}$ ($l^{\prime}_{\rm s}$) are the distances between the PRM (SRM) and BS along the $x$ ($y$) axis.  For aLIGO, $L_{\rm arm}=3995 \unit{m}$. All values are in units of m.}
\begin{ruledtabular}
\begin{tabular}{cccccccc}
 & $L_{\rm arm}$ & $l_{\rm x}$ & $l_{\rm y}$ & $l_{\rm p}$ & $l_{\rm s}$ & $l^{\prime}_{\rm p}$ & $l^{\prime}_{\rm s}$ \\
\hline
KAGRA & 3000 & 26.7 & 23.3 & 66.6 & 66.6 & 19.5 & 19.4 \\
\end{tabular}
\end{ruledtabular}
\end{table}

While the interferometer configuration is similar between KAGRA and aLIGO, the mirror substrate for the arm cavity test masses are different.
Where aLIGO employs room temperature fused silica mirrors for all the mirrors, KAGRA employs sapphire for cryogenic test masses, and fused silica for room temperature auxiliary mirrors. Cryogenic cooling of the test masses is a promising way to reduce thermal noise, and future gravitational wave detectors such as LIGO Voyager~\cite{LIGOVoyager}, Einstein Telescope~\cite{ET}, and Cosmic Explorer~\cite{CE} also plan to operate at cryogenic temperatures.

\begin{figure}
\begin{center}
\includegraphics[width=\hsize]{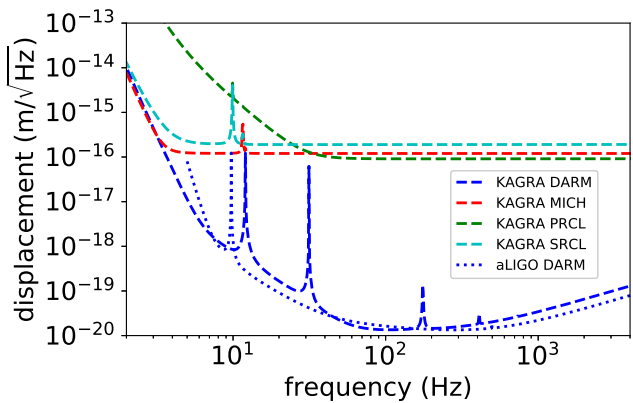}
\end{center}
\caption{\label{Displacement} The designed displacement sensitivity of KAGRA. The designed DARM sensitivity for aLIGO is also shown for comparison. Sensitivity data for KAGRA were obtained from \Reference{JGW-T2011755}, and that for aLIGO were obtained from \Reference{UpdatedaLIGODesign}.}
\end{figure}

The designed displacement sensitivity of KAGRA for each length degree of freedom is shown in \zu{Displacement}. For DARM, the sensitivity is limited by the seismic noise and the thermal noise at low frequencies, and the quantum noise at high frequencies, as described in detail in \Reference{PSOKAGRA}. For auxiliary degrees of freedom, the sensitivity is limited by the electronic noise in the mirror actuation below $\sim 4 \unit{Hz}$ for MICH and SRCL, below $\sim 40 \unit{Hz}$ for PRCL~\cite{KAGRAActuator}. The actuator noise for PRCL is larger because power recycling mirrors are suspended by simpler vibration isolation systems and require a larger actuation range. The sensitivity at higher frequencies is limited by quantum shot noise~\cite{AsoKAGRA}.

\section{Length changes from vector dark matter}
We focus on the massive vector field coupled with $B$ or $B-L$ current $J_D^\nu$, whose Lagrangian is given by
\begin{equation}
    \mathcal L 
    =
    -\frac{1}{4} F^{\mu\nu}F_{\mu\nu} 
    +\frac{1}{2}m_A^2 A^\nu A_\nu 
    - \epsilon_D e J_D^\nu A_\nu,
\end{equation}
where $F_{\mu\nu}\equiv \partial_\mu A_{\nu}-\partial_\nu A_\mu$
is the field strength, $m_A$ is the mass of the vector field, and
$\epsilon_D$ is the gauge coupling constant normalized to the electromagnetic coupling constant $e$.
The vector dark matter field at location $\vec{x}$ is given by
\begin{equation}
 \vec{A}(t,\vec{x}) = 
 \frac{\sqrt{2 \rho_{\rm DM}}}{m_A} \vec{e}_A 
 \cos{(m_A t - \vec{k} \cdot \vec{x} + \delta_{\tau}(t))} ,
\end{equation}
where $\vec{e}_A$ is the unit vector parallel to $\vec{A}$, $\rho_{\rm DM} \simeq 0.3 \unit{GeV/cm^3}$ is the local dark matter density, and $k=m_A v$ with $v \simeq 10^{-3}$ being the local velocity of dark matter. 
Note that $A_0$ is negligibly smaller than $A_i$, and hence we ignore it.
The phase factor, $\delta_\tau(t)$, can be regarded as constant within the coherent time scale $\tau=2 \pi/(mv^2)$.

In the same way as the electric force $\vec{F} = -qe\dot{\vec{A}}$ in electromagnetism,
the non-relativistic vector field accelerates a free-falling object $i$ with a charge $q_{D,i}$ and mass $M_i$ as
\begin{eqnarray}
 \vec{a}_i(t,\vec{x}_i) &=& 
 \epsilon_D e \frac{q_{D,i}}{M_i} \sqrt{2 \rho_{\rm DM}} 
 \vec{e}_A\sin{(m_A t - \vec{k} \cdot \vec{x}_i)} \\
 &\equiv& a_{i} 
 \vec{e}_A\sin{(m_A t - \vec{k} \cdot \vec{x}_i)} .
\end{eqnarray}
For a $U(1)_{B}$ gauge boson, $q_{B}/M = A/\mu/\mneutron \simeq 1/\mneutron$ is almost identical between different materials. Here, $A$ is the mass number, $\mu$ is the atomic mass in atomic units, and $\mneutron$ is the neutron mass. For fused silica and sapphire, $A/\mu-1$ is $5.5 \times 10^{-4}$ and $5.1 \times 10^{-4}$, respectively. On the other hand, for a $U(1)_{B-L}$ gauge boson, $q_{B-L}/M = (A-Z)/\mu/\mneutron$, where $Z$ is the atomic number, and the neutron ratio $(A-Z)/\mu \simeq 0.5$ would be more distinguishable between different materials. For fused silica and sapphire, the values are 0.501 and 0.51, respectively.

Let us consider two test masses placed along the $x$ axis, separated by a distance $L_{\rm x}$. The length changes along the $x$ axis from dark matter induced acceleration can be calculated by
\begin{equation}
\begin{split}
 \delta L_{\rm x} &= \frac{1}{2} \int {\rm d} t \int {\rm d} t \vec{e}_{x} \cdot [ \vec{a}_1(t,\vec{x}_1)  \\
 & \quad + \vec{a}_1(t-2 L_{\rm x},\vec{x}_1) - 2 \vec{a}_2(t-L_{\rm x},\vec{x}_2) ] ,
\end{split}
\end{equation}
where $\vec{e}_{x}$ is the unit vector along the $x$ axis. For $m_A L_{\rm x} \ll 1$, which is the case in the mass range KAGRA and aLIGO can probe, the amplitude of the oscillating length change is given by
\begin{equation}
\begin{split}
 \delta L_{{\rm x},0} &= \frac{\cos \Omega_A}{m_A^2} \left[ \left(a_1\left(1-\frac{m_A^2 L_{\rm x}^2}{2} \right)-a_2\right)^2 \right. \\
 & \quad \left. + a_1a_2 (kL_{\rm x} \cos \Omega_k)^2 \right]^{1/2} ,
\end{split}
\end{equation}
where $\Omega_A$ is the angle between $\vec{A}$ and the $x$ axis, and $\Omega_k$ is the angle between $\vec{k}$ and the $x$ axis. At around 100~Hz, where ground-based gravitational wave detectors are most sensitive, $m_A \simeq 4 \times 10^{-13} \unit{eV}$, and $2 \pi / k \simeq 3 \times 10^9 \unit{m}$. Therefore, $m_A^2 L_{\rm x}^2$ and $kL_{\rm x}$ are on the order of $10^{-5}$ for kilometer-scale interferometers, and the length changes are largely attenuated when $a_1 = a_2$. In the case where $a_1 = a_2$, the $m_A^2 L_{\rm x}^2$ term, which comes from the finite light travel time between two test masses, dominates in the mass range $m_A \gg v/L_{\rm x}$~\cite{MorisakiInPrep}.

By taking the average over all possible directions of $\vec{A}$ and $\vec{k}$, we obtain
\begin{equation}
\begin{split}
 \sqrt{\langle \delta L_{{\rm x},0}^2 \rangle} &= \frac{1}{3 m_A^2} \left[ 3 \left(a_1\left(1-\frac{m_A^2 L_{\rm x}^2}{2} \right)-a_2\right)^2 \right. \\
 & \quad \left. + a_1a_2  (kL_{\rm x})^2 \right]^{1/2} . \label{eq:average}
\end{split}
\end{equation}
For simplicity, we assumed that the directions of $\vec{A}$ and $\vec{k}$ are not correlated, as the three components of $\vec{A}$ are in equilibrium, although vector dark matter might have only had its longitudinal or transverse modes in the early universe, depending on the production mechanism~\cite{GrahamInflation2016,Dror2019,Co2019,Bastero-Gil2019,Agrawal2020,Nakayama2019,Nakai2020}.
A similar calculation can be performed for two mirrors along the $y$ axis, and we can obtain the same result for the angular average.

By substituting the above equations into Eqs.~(\ref{eq:darm})--(\ref{eq:srcl}), the average amplitude of the oscillating length changes from the vector field can be written using the lengths in \hyou{Lengths} as
\begin{eqnarray}
 \sqrt{\langle \delta L_{{\rm DARM},0}^2 \rangle} &=& \sqrt{\langle \delta L_{{\rm x},0}^2 \rangle + \langle \delta L_{{\rm y},0}^2 \rangle} \\
 \sqrt{\langle \delta L_{{\rm MICH},0}^2 \rangle} &=& \sqrt{\langle \delta l_{{\rm x},0}^2 \rangle + \langle \delta l_{{\rm y},0}^2 \rangle} \\
 \sqrt{\langle \delta L_{{\rm PRCL},0}^2 \rangle} &=& \frac{\sqrt{\langle (\delta l_{{\rm x},0} +2 \delta l_{{\rm p},0}^\prime)^2 \rangle + \langle \delta l_{{\rm y},0}^2 \rangle}}{2} \\
 \sqrt{\langle \delta L_{{\rm SRCL},0}^2 \rangle} &=& \frac{\sqrt{\langle \delta l_{{\rm x},0}^2 \rangle + \langle (\delta l_{{\rm y},0}+2 \delta l_{{\rm s},0}^\prime)^2 \rangle}}{2} .
\end{eqnarray}
The effect of the folding mirrors of the recycling cavities is negligible, since two folding mirrors are made of the same material and the distance between the folding mirrors is 10--20~m for both KAGRA and aLIGO. Note that the angular average can be computed independently for the lengths along the $x$ axis and $y$ axis, since the two axes are orthogonal.

\begin{figure}
\begin{center}
\includegraphics[width=\hsize]{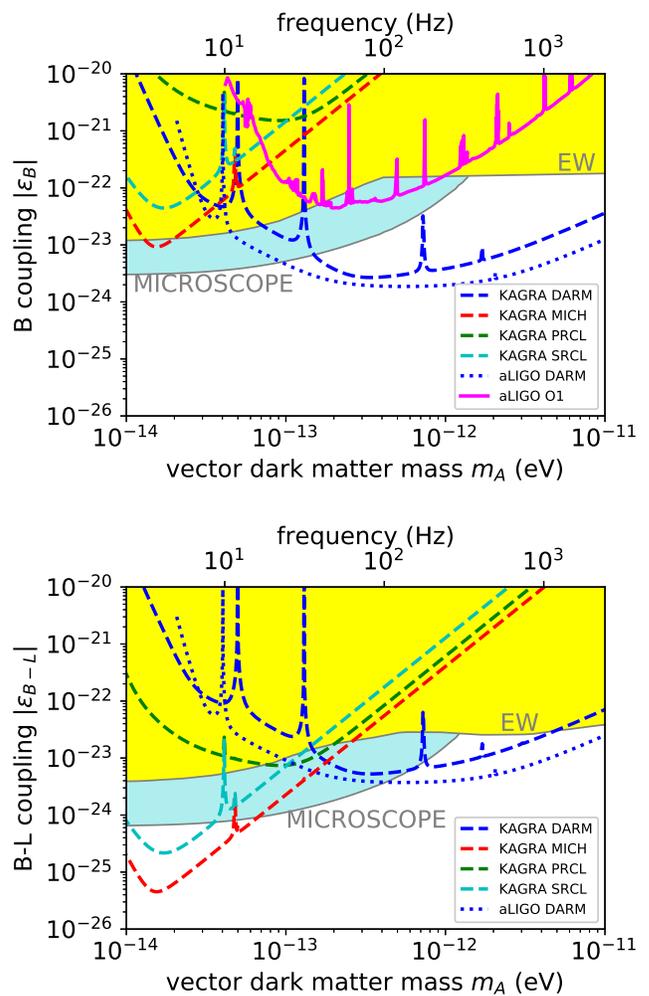}
\end{center}
\caption{\label{Limits} The projected sensitivity for $B$ (upper) and $B-L$ (lower) vector dark matter couplings of KAGRA and aLIGO with the measurement time of $T_{\rm obs}=1$~year. The shaded regions show bounds from fifth-force searches with E{\"o}t-Wash torsion pendulum~\cite{EW2008,EW2012} (yellow) and MICROSCOPE satellite~\cite{MICROSCOPE2018} (light blue). Bounds on the coupling constant of the Yukawa potential $\alpha$ from these fifth-force searches are converted to $\epsilon_D$ using $\epsilon_D^2 = \alpha G \mneutron^2/(e^2/4 \pi)$, where $G$ is the gravitational constant. The bound for $B$ coupling from 893 hours of aLIGO O1 DARM data~\cite{Guo2019} is also shown for comparison.}
\end{figure}

\section{Sensitivity to $B$ and $B-L$ coupling}
With a one-sided spectrum of the displacement sensitivity of $\sqrt{S_{d}(m_A)}$, the signal-to-noise ratio (SNR) to the length changes from the vector field on average $\sqrt{\langle \delta L_{d,0}^2 \rangle}$ is given by
\begin{equation}
 {\rm SNR} = \frac{\sqrt{T_{\rm eff}}}{2 \sqrt{S_{d}(m_A)}} \sqrt{\langle \delta L_{d,0}^2 \rangle} ,
\end{equation}
where $d$ runs from DARM, MICH, PRCL and SRCL. When the total measurement time $T_{\rm obs}$ is shorter than the coherent time $\tau$, the dark matter field oscillation can be regarded as coherent and $T_{\rm eff} = T_{\rm obs}$. However, when $T_{\rm obs} \gtrsim \tau$, the phase factor $\delta_\tau$ behaves as a random variable remaining constant for each period of $\tau$, and $T_{\rm eff}=\sqrt{T_{\rm obs} \tau}$~\cite{CASPEr}. At 100~Hz or $m_A \simeq 4 \times 10^{-13} \unit{eV}$, the coherent time is $\tau \simeq 10^4 \unit{sec}$.

By setting the SNR to unity, we obtain the detectable limit to $\sqrt{\langle \delta L_{d,0}^2 \rangle}$, which can be converted into the coupling constant $\epsilon_D$ using the equations above. Figure~\ref{Limits} shows the projected sensitivity of KAGRA and aLIGO for $B$ and $B-L$ vector dark matter couplings. Here, the displacement sensitivity shown in \zu{Displacement} and length parameters in \hyou{Lengths} are used, and $T_{\rm obs}$ is set to 1 year.

For $B$ coupling, DARM gives the best sensitivity for almost the entire mass range, owing to the longest interferometer length. Below $m_{A} \lesssim 7 \times 10^{-14}$~eV, the sensitivity from the $kL$ term in \shiki{eq:average} dominates, while above that mass, the sensitivity from the $m_A^2 L^2$ term dominates. For $m_{A} \lesssim 4 \times 10^{-14}$~eV, auxiliary channels provide better sensitivity, which is due to the difference in the charge density $q_{B}/M$ between sapphire and fused silica at the $10^{-5}$ level. On the other hand, for $B-L$ coupling, auxiliary channels provide better sensitivity for $m_{A} \lesssim 2 \times 10^{-13}$~eV than DARM, owing to the difference in $q_{B-L}/M$ between sapphire and fused silica at the $10^{-2}$ level. In particular, when the MICH channel is used, KAGRA can reach $\epsilon_{B-L} \simeq 4.5 \times 10^{-26}$ at $m_{A} \simeq 1.5 \times 10^{-14}$~eV, which is more than an order of magnitude improvement compared with the bounds set by the MICROSCOPE experiment~\cite{MICROSCOPE2018}.

We note that there are various technical noises at low frequencies that could degrade the sensitivity for vector dark matter in a lower mass range. In aLIGO during its first observing run O1, the displacement sensitivity on the order of $10^{-16} \unit{m/\sqrt{Hz}}$ was achieved above $\sim 20$~Hz for auxiliary length channels~\cite{aLIGOO1NB}. At lower frequencies, the beam splitter motion and electronic noises in the local sensors for the damping control of the suspension resonances were larger than the seismic noise and actuator noise. Such noises can be reduced by feed-forward cancellation techniques and improved local sensors~\cite{LIGO-LF}.

It is worth pointing out that our scheme can be applied to KAGRA without any modification to the existing interferometer and without losing any sensitivity to gravitational waves. The use of auxiliary channels has an advantage over the main DARM channel in which it is easier to differentiate dark matter signals from gravitational wave signals. This is because the length change caused by gravitational waves is smaller than that for DARM by two orders of magnitude, owing to the shorter interferometer length. The sensitivity for the $B-L$ vector dark matter similar to KAGRA can also be achieved with aLIGO if the auxiliary mirrors are replaced with, for example, sapphire mirrors. Changing the material of the auxiliary mirrors has negligible impact on the gravitational wave sensitivity, since the coupling of the displacement noise of the auxiliary mirrors to DARM is smaller than $\mathcal{O}(10^{-2})$ in the observation band~\cite{aLIGOO1NB}.

\section{Conclusion}
We proposed the use of auxiliary length channels from gravitational wave detectors to search for ultralight vector dark matter. We have shown that when our scheme is applied to $B-L$ vector dark matter search with the cryogenic gravitational wave detector KAGRA, the sensitivity for $m_{A} \lesssim 2 \times 10^{-13}$~eV can be improved, compared with the search using the main differential arm length channel and other fifth-force experiments. It is shown that more than an order of magnitude of the unexplored region can be probed at masses around $1.5 \times 10^{-14}$~eV. This is because KAGRA uses different substrates for the test masses and auxiliary mirrors. The auxiliary channels measure the changes in the lengths between the sapphire test masses and fused silica auxiliary mirrors, which have different $B-L$ charge densities.

Future gravitational wave detectors also plan to use cryogenic arm cavity test masses that are made of different substrates from that of room temperature auxiliary mirrors. The sensitivity to ultralight dark matter depends on the detailed design of the auxiliary mirrors, but the sensitivity improvement over the designed sensitivity of KAGRA can be expected, not only for gravitational waves, but also for signals from ultralight dark matter. Further sensitivity enhancement can be done by optimizing the design of the auxiliary mirrors, without reducing any the sensitivity to gravitational waves. Our study is the first proposal to focus on auxiliary length channels to use them as science data. Our proposal opens up new possibilities for dark matter searches using gravitational wave detectors.

\begin{acknowledgments}
We thank Masahiro Ibe, Kentaro Komori, Koji Nagano, Yutaro Enomoto and Denis Martynov for invaluable inputs and stimulating discussions. This work was supported by JSPS KAKENHI Grant Numbers 18H01224, 18K13537, 18K18763, 19J13840, 19J21974, and JST CREST Grant Number JPMJCR1873. H.N. is supported by the Advanced Leading Graduate Course for Photon Science, and I.O. is supported by the JSPS Overseas Research Fellowship.
\end{acknowledgments}

%


\begin{thebibliography}{99}
\bibitem{NewEra}
G.~Bertone and T.~M.~P.~Tait,
Nature \href{https://doi.org/10.1038/s41586-018-0542-z}{{\bf 562}, 51 (2018)}.

\bibitem{ULDMWitten}
L.~Hui, J.~P.~Ostriker, S.~Tremaine, and E.~Witten,
Phys. Rev. D \href{https://doi.org/10.1103/PhysRevD.95.043541}{{\bf 95}, 043541 (2017)}.

\bibitem{ULDMFerreira}E
E.~G.~M.~Ferreira, 
\href{https://arxiv.org/abs/2005.03254}{arXiv:2005.03254}.

\bibitem{AxionInterferometry}
W.~DeRocco and A.~Hook,
Phys. Rev. D \href{https://doi.org/10.1103/PhysRevD.98.035021}{{\bf 98}, 035021 (2018)}.

\bibitem{DANCE}
I.~Obata, T.~Fujita, and Y.~Michimura,
Phys. Rev. Lett. \href{https://doi.org/10.1103/PhysRevLett.121.161301}{{\bf 121}, 161301 (2018)}.

\bibitem{ADBC}
H.~Liu, B.~D.~Elwood, M.~Evans, and J.~Thaler,
Phys. Rev. D \href{https://doi.org/10.1103/PhysRevD.100.023548}{{\bf 100}, 023548 (2019)}.

\bibitem{LinearAxion}
K.~Nagano, T.~Fujita, Y.~Michimura, and I.~Obata,
Phys. Rev. Lett. \href{https://doi.org/10.1103/PhysRevLett.123.111301}{{\bf 123}, 111301 (2019)}.

\bibitem{QuantumAxion}
D. Martynov and H. Miao,
Phys. Rev. D \href{https://doi.org/10.1103/PhysRevD.101.095034}{{\bf 101}, 095034 (2020)}.

\bibitem{Stadnik2015}
Y.~V.~Stadnik and V.~V.~Flambaum,
Phys. Rev. Lett. \href{https://doi.org/10.1103/PhysRevLett.114.161301}{{\bf 114}, 161301 (2015)}.

\bibitem{Stadnik2016}
Y.~V.~Stadnik and V.~V.~Flambaum,
Phys. Rev. A \href{https://doi.org/10.1103/PhysRevA.93.063630}{{\bf 93}, 063630 (2016)}.

\bibitem{Geraci2019}
A.~A.~Geraci, C.~Bradley, D.~Gao, J.~Weinstein, and A.~Derevianko,
Phys. Rev. Lett. \href{https://doi.org/10.1103/PhysRevLett.123.031304}{{\bf 123}, 031304 (2019)}.

\bibitem{Grote2019}
H.~Grote and Y.~V.~Stadnik,
Phys. Rev. Research \href{https://doi.org/10.1103/PhysRevResearch.1.033187}{{\bf 1}, 033187 (2019)}.

\bibitem{Arvanitaki2015}
A.~Arvanitaki, J.~Huang, and K.~Van Tilburg,
Phys. Rev. D \href{https://doi.org/10.1103/PhysRevD.91.015015}{{\bf 91}, 015015 (2015)}.

\bibitem{Morisaki2019}
S.~Morisaki and T.~Suyama,
Phys. Rev. D \href{https://doi.org/10.1103/PhysRevD.100.123512}{{\bf 100}, 123512 (2019)}.

\bibitem{Graham2016}
P.~W.~Graham, D.~E.~Kaplan, J.~Mardon, S.~Rajendran, and W.~A.~Terrano,
Phys. Rev. D \href{https://doi.org/10.1103/PhysRevD.93.075029}{{\bf 93}, 075029 (2016)}.

\bibitem{Pierce2018}
A.~Pierce, K.~Riles, and Y.~Zhao,
Phys. Rev. Lett. \href{https://doi.org/10.1103/PhysRevLett.121.061102}{{\bf 121}, 061102 (2018)}.

\bibitem{Carney2019}
D.~Carney, A.~Hook, Z.~Liu, J.~M.~Taylor, and Y.~Zhao,
\href{https://arxiv.org/abs/1908.04797}{arXiv:1908.04797}.

\bibitem{Manley2020}
J.~Manley, M.~D.~Chowdhury, D.~Grin, S.~Singh, and D.~J.~Wilson, \href{https://arxiv.org/abs/2007.04899}{arXiv:2007.04899}.

\bibitem{aLIGO}
J.~Aasi \etal (The LIGO Scientific Collaboration),
Classical Quantum Gravity \href{https://doi.org/10.1088/0264-9381/32/7/074001}{{\bf 32}, 074001 (2015)}.

\bibitem{aLIGOO1NB}
D.~V.~Martynov, E.~D.~Hall \etal,
Phys. Rev. D \href{https://doi.org/10.1103/PhysRevD.93.112004}{{\bf 93}, 112004 (2016)}.

\bibitem{AdV}
F.~Acernese \etal (Virgo Collaboration),
Classical Quantum Gravity \href{https://doi.org/10.1088/0264-9381/32/2/024001}{{\bf 32}, 024001 (2015)}.

\bibitem{AsoKAGRA}
Y.~Aso, Y.~Michimura, K.~Somiya, M.~Ando, O.~Miyakawa, T.~Sekiguchi, D.~Tatsumi, and H.~Yamamoto (The KAGRA Collaboration),
Phys. Rev. D \href{https://doi.org/10.1103/PhysRevD.88.043007}{{\bf 88}, 043007 (2013)}.

\bibitem{bKAGRAPhase1}
T.~Akutsu \etal (KAGRA Collaboration),
Classical Quantum Gravity \href{https://dx.doi.org/10.1088/1361-6382/ab28a9}{{\bf 36}, 165008 (2019)}.

\bibitem{Green:1984sg}
M.~B.~Green and J.~H.~Schwarz,
Phys. Lett. B \href{https://doi.org/10.1016/0370-2693(84)91565-X}{{\bf 149}, 117 (1984)}.


\bibitem{LIGOVoyager}
R.~X.~Adhikari \etal,
Classical Quantum Gravity \href{https://doi.org/10.1088/1361-6382/ab9143}{{\bf 37}, 165003 (2020)}.

\bibitem{ET}
M.~Punturo \etal,
Classical Quantum Gravity \href{https://doi.org/10.1088/0264-9381/27/19/194002}{{\bf 27}, 194002 (2010)}.

\bibitem{CE}
B.~P.~Abbott \etal (LIGO Scientific Collaboration),
Classical Quantum Gravity \href{https://doi.org/10.1088/1361-6382/aa51f4}{{\bf 34}, 044001 (2017)}.

\bibitem{JGW-T2011755}
Y.~Michimura, K.~Komori \etal,
{\it Estimated sensitivity for auxiliary degrees of freedom of the KAGRA interferometer},
JGW Document No. \href{https://gwdoc.icrr.u-tokyo.ac.jp/cgi-bin/DocDB/ShowDocument?docid=11755}{JGW-T2011755} (2020),
{\tt https://gwdoc.icrr.u-tokyo.ac.jp\slash{}cgi-bin\slash{}DocDB\slash{}ShowDocument?docid=11755}.

\bibitem{UpdatedaLIGODesign}
L.~Barsotti, S.~Gras, M.~Evans, and P.~Fritschel,
{\it Updated Advanced LIGO sensitivity design curve},
LIGO Report No. \href{https://dcc.ligo.org/LIGO-T1800044/public}{LIGO-T1800044} (2018),
{\tt https://dcc.ligo.org\slash{}{LIGO-T1800044}\slash{}public}.

\bibitem{PSOKAGRA}
Y.~Michimura, K.~Komori, A.~Nishizawa, H.~Takeda, K.~Nagano, Y.~Enomoto, K.~Hayama, K.~Somiya, and M.~Ando,
Phys. Rev. D \href{https://doi.org/10.1103/PhysRevD.97.122003}{{\bf 97}, 122003 (2018)}.

\bibitem{KAGRAActuator}
Y.~Michimura \etal,
Classical Quantum Gravity \href{https://doi.org/10.1088/1361-6382/aa90e3}{{\bf 34}, 225001 (2017)}.

\bibitem{Nelson2011}
A.~E.~Nelson and J.~Scholtz,
Phys. Rev. D \href{https://doi.org/10.1103/PhysRevD.84.103501}{{\bf 84}, 103501 (2011)}.

\bibitem{GrahamInflation2016}
P.~W.~Graham, J.~Mardon, and S.~Rajendran,
Phys. Rev. D \href{https://doi.org/10.1103/PhysRevD.93.103520}{{\bf 93}, 103520 (2016)}.

\bibitem{Dror2019}
J.~A.~Dror, K.~Harigaya, and V.~Narayan,
Phys. Rev. D \href{https://doi.org/10.1103/PhysRevD.99.035036}{{\bf 99}, 035036 (2019)}.

\bibitem{Co2019}
R.~T.~Co, A.~Pierce, Z.~Zhang, Y.~Zhao,
Phys. Rev. D \href{https://doi.org/10.1103/PhysRevD.99.075002}{{\bf 99}, 075002 (2019)}.

\bibitem{Bastero-Gil2019}
M.~Bastero-Gil, J.~Santiago, L.~Ubaldi, and R.~Vega-Morales, 
J. Cosmol. Astropart. Phys. \href{https://doi.org/10.1088/1475-7516/2019/04/015}{{\bf 04}, 015 (2019)}.

\bibitem{Agrawal2020}
P.~Agrawal, N.~Kitajima, M.~Reece, T.~Sekiguchi, F.~Takahashi,
Phys. Lett. B \href{https://doi.org/10.1016/j.physletb.2019.135136}{{\bf 801}, 135136 (2020)}.

\bibitem{Nakayama2019}
K.~Nakayama,
J. Cosmol. Astropart. Phys. \href{https://doi.org/10.1088/1475-7516/2019/10/019}{{\bf 10}, 019 (2019)}.

\bibitem{Nakai2020}
Y.~Nakai, R.~Namba, Z.~Wang,
\href{https://arxiv.org/abs/2004.10743}{arXiv:2004.10743}.

\bibitem{MorisakiInPrep}
S.~Morisaki \etal, {\it in preparation}.

\bibitem{EW2008}
S.~Schlamminger, K.-Y.~Choi, T.~A.~Wagner, J.~H.~Gundlach, and E.~G.~Adelberger,
Phys. Rev. Lett. \href{https://doi.org/10.1103/PhysRevLett.100.041101}{{\bf 100}, 041101 (2008)}.

\bibitem{EW2012}
T.~A.~Wagner, S.~Schlamminger, J.~H.~Gundlach, and E.~G.~Adelberger,
Classical Quantum Gravity \href{https://doi.org/10.1088/0264-9381/29/18/184002}{{\bf 29}, 184002 (2012)}.

\bibitem{MICROSCOPE2018}
J.~Berg{\'e}, P.~Brax, G.~M{\'e}tris, M.~Pernot-Borr{\`a}s, P.~Touboul, and J.-P.~Uzan,
Phys. Rev. Lett. \href{https://doi.org/10.1103/PhysRevLett.120.141101}{{\bf 120}, 141101 (2018)}.

\bibitem{Guo2019}
H.-K.~Guo, K.~Riles, F.-W.~Yang, Y.~Zhao,
Communications Physics \href{https://doi.org/10.1038/s42005-019-0255-0}{{\bf 2}, 155 (2019)}.

\bibitem{CASPEr}
D.~Budker, P.~W.~Graham, M.~Ledbetter, S.~Rajendran, and A.~O.~Sushkov,
Phys. Rev. X \href{https://doi.org/10.1103/PhysRevX.4.021030}{{\bf 4}, 021030 (2014)}.

\bibitem{LIGO-LF}
H.~Yu, D.~Martynov, S.~Vitale, M.~Evans, D.~Shoemaker, B.~Barr, G.~Hammond, S.~Hild, J.~Hough, S.~Huttner, S.~Rowan, B.~Sorazu, L.~Carbone, A.~Freise, C.~Mow-Lowry, K.~L.~Dooley, P.~Fulda, H.~Grote, and D.~Sigg,
Phys. Rev. Lett. \href{https://doi.org/10.1103/PhysRevLett.120.141102}{{\bf 120}, 141102 (2018)}.

\end{thebibliography}
\end{document}